# Taking Quantitative Genomics into the Wild.

Susan E. Johnston[1], Nancy Chen[2], Emily B. Josephs[3]

[1] Institute of Ecology and Evolution, School of Biological Sciences, University of Edinburgh, Edinburgh, UK.

[2] Department of Biology, University of Rochester, Rochester, New York, USA.

[3] Department of Plant Biology and Ecology, Evolution, and Behavior Program, Michigan State University, East Lansing, MI, USA.

## 1. Introduction.

A key goal in studies of ecology and evolution is understanding the causes of phenotypic diversity in nature. Most traits of interest, such as those relating to morphology, **life-history**, immunity and behaviour are quantitative, and phenotypic variation is driven by the cumulative effects of genetic and environmental variation [1]. The field of **quantitative genetics** aims to quantify the additive genetic component of this trait variance (i.e. the "**heritability**"), often with the underlying assumption that trait variance is driven by many loci of very tiny ("infinitesimal") effects throughout the genome [2,3]. This approach allows us to understand the evolutionary potential of natural populations and can be extended to examine the genetic covariation with **fitness** to predict responses to selection [4]. Therefore, quantitative genetic studies are fundamental to understanding evolution in the wild [1].

Over the last two decades, there has been a wealth of quantitative genetic studies investigating trait heritabilities and **genetic correlations**, but these studies were initially limited to long-term studies of pedigreed populations, or common-garden experiments in organisms more easily suited to crossing and rearing in a controlled environment. However, the advent of tractable genomic technologies has since allowed quantitative genetic studies to be conducted in a more diverse range of wild systems and has increased the opportunities for addressing outstanding questions in ecology and evolution. In particular, genomic studies can uncover the genetic basis of fitness-related quantitative traits, allowing a better understanding of their evolutionary dynamics.

We organised this special issue to highlight new work and review recent advances at the cutting edge of "Wild Quantitative Genomics". In this Editorial, we will present some history of wild quantitative genetic and genomic studies, before discussing the main themes in the papers published in this special issue and highlighting the future outlook of this dynamic field.

## 2. A Short History of Quantitative Genetics in the Wild.

### (a) Traditional quantitative genetics.

Quantitative genetic models have traditionally estimated the heritability of traits using information on the pairwise relationships between individuals within a population [3]. These models, such as the commonly used "animal model" [5], were originally developed for animal and plant breeding experiments with experimentally-determined relatedness and have become powerful tools in predicting responses to artificial selection [6]. From the 1980s onwards, DNA parentage analysis gradually became more tractable in non-model species, providing the opportunity to construct pedigrees and thus quantify pairwise relatedness in previously well-studied populations [7–9]. These new analyses allowed quantitative genetics to address a range of outstanding questions in wild systems living in natural conditions, including the heritability of life-history traits [10], the contribution of common environment effects to phenotypic variation [11], and the paradox of evolutionary stasis, where traits do not evolve despite apparent selection [12]. However, these traditional models cannot be refined beyond the assumption that traits are controlled by an infinite number of genes of infinitely small effect (known as the "infinitesimal model") [13]. In addition, traditional models only consider the expected pedigree relatedness between individuals, whereas the true relatedness between individuals can deviate from this expectation due to the randomness of recombination and Mendelian segregation [14]. More critically, such studies were almost always restricted to established long-term studies of mammals and birds, where multiple generations of individuals were already available for pedigree construction. Whilst these long-term, individual-based studies have been crucial in establishing a solid foundation for what is feasible in wild populations, they have tended to be biassed towards smaller populations and organisms that do not move long distances, making them easier to study and sample. Gienapp and colleagues [4] referred to this as a "tool of privilege", as it remained challenging to establish and fund long-term quantitative genetic studies in new and more diverse study systems.

Alongside studies of pedigreed populations in the wild, studies of plants and other organisms that could be reared in controlled conditions have investigated quantitative genetics using common garden studies [15]. These studies often use crossing designs to generate populations with known relatedness between individuals. Then, these individuals are phenotyped in a common field environment to quantify genetic variation for phenotypes of interest. These powerful approaches have allowed quantitative geneticists to estimate additive genetic variance for fitness and evaluate the potential for adaptation to changing environments [16], measure **inbreeding depression** [17], and determine the mechanisms of genetic correlations [18], amongst other questions. However, as with pedigree-based approaches discussed above, common garden studies have some limitations. First, common garden studies require a design where specific individuals can be marked so that they are distinguishable from each other. Second, these experiments generally study individuals from emergence to senescence, missing key parts of the life cycle [19,20]. Finally, these studies are often only able to look at one or a few years, limiting their ability to describe selection that may vary across years [21].

### (b) The genomics era.

The rise of high-throughput DNA sequencing and genotyping arrays created an unprecedented opportunity to bring quantitative genetics to a much wider range of taxa and research directions [4,14]. The characterisation of individual genomic variation, most often through single nucleotide polymorphisms (SNPs), has led to two key advances that genomics offers over traditional methods. First, it is now possible to quantify the actual proportion of genomic variation shared between individuals in wild populations (i.e. genomic-pairwise rather than pedigree-pairwise relationships) [22–27]. This is a major development in the field, and means that quantitative genetic studies could feasibly be carried out in any system with a suitable number of phenotyped individuals, even without pedigrees, provided there are enough genomic markers and related individuals sampled (i.e. that markers are in **linkage disequilibrium** with causal variants [4,24,25,28]). Second, it is increasingly possible to decompose heritable variation into its underlying **genetic architecture** – that is, use genetic mapping techniques to discover the number, effect sizes, distribution, and interactions of the underlying genes throughout the genome [29,30] (but see [31]), as well as the contribution of **epigenetic** mechanisms [32] and **indirect genetic effects**, such as maternal genetic effects [25,33,34].

### 3. Wild Quantitative Genomics Today.

These advances have created opportunities to better understand selection and evolution in wild populations. In particular, they have provided a foundation for detailed dissection of the genomic architecture of additive genetic variation [27,32,35], local adaptation [36,37], inbreeding depression [38,39], indirect genetic effects and multi-genomic architectures [25,40], genotype-by-environment interactions [41] and **multivariate** genetic architectures and selection [42–44], as well as more accurate predictions of evolutionary responses and constraints [45–47] and the genetic pathways important to local adaptation and fitness [48]. In this special issue, we present twelve contributed publications[1] that encompass broad themes in cutting-edge wild quantitative genomics.

#### (a) Understanding the genetic architecture of quantitative traits.

A common goal of quantitative genomic studies is determining the genetic architecture of traits. In this issue, Kelly [47] notes that this allows us to peer into the "black box" of quantitative genetic models and to see the genetic mechanisms shaping trait variation. A first step is to quantify the trait heritability; this measure can indicate the **adaptive potential** of populations, and heritable variation is necessary to identify causal loci. Duntsch *et al.* [27] quantify the heritability of morphological traits in the hihi (*Notiomytis cincta*, also known as the stitchbird), a passerine bird endemic to Aotearoa | New Zealand with two approaches: a traditional pedigree relatedness matrix between individuals, and a **genomic relatedness matrix (GRM)** quantifying the realised relatedness between individuals based on dense SNP genotype information. The similarity of their findings between pedigree and GRM approaches provides useful evidence that quantitative genomic studies do not require pedigrees to estimate trait heritability, provided there are enough related individuals and SNPs to capture the genetic variation within the population [24].

---

[1] For review purposes, PDF copies of the 12 submissions can be accessed here: https://www.dropbox.com/sh/35pdiux4hcas2tw/AACWKX7x9wJ59iEmkS2__DFVa?dl=0

A second step is determining the genomic regions contributing to heritable variation. **Quantitative trait locus (QTL)** mapping is an established approach that is still used to great effectiveness in well-phenotyped organisms where large numbers of molecular markers can be generated. Two studies in the issue used genotyping-by-sequencing data to identify QTLs; Lin *et al.* [36] used an F$_3$ mapping population to map several fecundity QTLs in *Boechera stricta* (Drummond's rockcress), and Coughlan *et al.* [37] used an F$_2$ mapping population from monkeyflowers *Mimulus guttatus* and *M. decorus* to identify 12 QTLs in 6 independent genomic regions associated with life-history traits important in local adaptation. The success of QTL mapping in controlled family crosses relies on generating co-inheritance of genetic markers and causal loci in subsequent generations. By contrast, **genome-wide association studies (GWAS)** leverage natural genetic variation and linkage disequilibrium patterns already present in the population to map major-effect loci to a high resolution. In this issue, Tietgen *et al.* [35] show the ability to map a large-effect locus in a free-living population using GWAS using a dense SNP array, identifying *MC1R* as a major locus controlling coat colour in Arctic foxes (*Vulpes lagopus*).

One challenge for genotype-phenotype association studies is that traditional QTL and GWAS approaches can only detect moderate to large-effect loci, limiting our ability to study traits with **polygenic** architectures (i.e. controlled by many loci of small effects) and leading to a bias towards investigating traits with a simple genetic basis [31,49]. However, this special issue presents several approaches to characterise the nature of polygenic variation. For example, Duntsch *et al.* [27] did not identify significant loci in a GWAS but were able to quantify the contributions of individual chromosomes to the additive genetic variance using a technique called "chromosome partitioning." They showed that the contribution of a chromosome to additive genetic variance was proportional to its length, indicating a polygenic architecture. Another approach to characterising polygenic variation and adaptation – reviewed by McGaugh *et al.* [46] – is **genomic prediction**, where all SNP marker effects are estimated in the same model with a mixture of effect size distributions, in turn directly modelling how the whole genome affects a given trait [50]. This approach was used in a precursor study to that of Hunter *et al.* [45] to quantify the heritability and trait architectures of morphological traits in a wild population of Soay sheep (*Ovis aries*) [51]. We describe in more detail the utility of genomic prediction for studies of evolutionary change below.

A final point on the genetic architecture of traits is addressed in a review by Husby [32], who discusses the potential contribution of epigenetic variation to phenotypic variation in wild populations, including from small RNAs, DNA methylation, and histone modifications. Epigenetic mechanisms have been shown to contribute to variation in developmental, morphological, and reproductive traits in diverse species [52,53], although there is some suggestion that this contribution may be small and is likely to be in conjunction with genetic mechanisms [54]. Husby presents a number of tractable methods for studying epigenetic variation in wild populations, but also highlights several unanswered questions that could be addressed in future, such as the strength of selection, heritability, and relative contribution to phenotypic variance of epigenetic markers, by combining approaches such as GWAS and epigenome-wide association studies (EWAS; the association of epigenetic marks with phenotype) into quantitative genetic models.

### (b) Linking genotypic variation with local adaptation and fitness.

Several papers in this issue tackle the most evolutionarily important of traits, fitness. Fitness is unique among quantitative traits because it is what is "seen" by natural selection [55]. When the genetic architecture of fitness or fitness-related traits is known, we can understand the evolutionary forces maintaining observed variation and/or their potential role in local adaptation. The study by Lin *et al.* in several populations of *Boechera stricta* identified multiple QTLs for traits directly related to fecundity, including a major effect QTL associated with higher and sturdier flowering stalks [36]. Variation at this QTL was associated with water availability across the landscape, and other fecundity QTLs showed high differentiation between genetic groups. These findings, along with significant differences in selection coefficients at different locations, are consistent with local adaptation maintaining variation in fitness.

Determining the genetic architecture of fitness-related traits can also help us understand why phenotypic variation persists. The study mapping coat colour variation in Arctic foxes [35] showed that heterozygous blue arctic foxes have higher fitness than homozygous white counterparts (comprising 90% of the population) and that this difference is likely mediated via an association with prey availability rather than other adaptations to winter, such as thermoregulation and camouflage. These results provide an explanation for the maintenance of variation of coat colour within this species.

**(c) Quantifying and predicting evolutionary change across the genome.**

Several papers in this issue present novel approaches to investigate selection and evolutionary change using information from all markers throughout the genome. One of the most promising new directions is the use of genomic prediction models [50]. This approach is well-established in animal and plant breeding and uses information from all SNPs in the genome to determine an individual's genomic estimated breeding values (GEBVs) for a trait of interest, without requiring pedigree information. A review by McGaugh *et al.* [46] describes how genomic prediction approaches open up new research avenues in wild populations, including: (i) determining the genetic architecture of traits (see 2(a) above); (ii) estimating GEBVs in unphenotyped individuals and historical samples; (iii) incorporating information on the error around GEBV estimates, overcoming previous criticisms of pedigree-based approaches (see [56]); (iv) choosing which individuals to breed, either in studies of mate choice or for conservation purposes; and (v) micro-evolutionary studies investigating changes in GEBVs over time. This last research question is addressed empirically by Hunter *et al.* [45], who investigate micro-evolutionary change in body weight in Soay sheep over a 35-year period. They show that despite a decrease in adult body weight over this period, there has been an increase in GEBVs at a rate greater than expected due to genetic drift, indicating that cryptic selection for increased body weight has occurred. This can arise when a trait has changed in response to selection, but a change in environment acts in the opposite direction on phenotype, masking changes in the underlying genotypes [57].

Another approach to investigate selection using genomic information is presented by Kelly [47]. His paper advocates the use of selection component analyses (SCA) to estimate selection at specific life-cycle stages by relating individual genotypes to fitness components. For example, viability and sexual selection can be measured by comparing allele frequencies in surviving and non-surviving individuals, or reproducing and non-reproducing individuals, respectively. The paper uses simulation approaches to demonstrate the effectiveness of this method in field experiments of *Mimulus* *guttatus* and *Drosophila melanogaster*, and provides practical advice on study design, particularly in the use of imprecise genotyping via low-coverage sequencing of large numbers of individuals to improve the power of this method.

Stahlke *et al.* [48] also use a population genetics approach to detect evolution in natural populations, investigating allele frequency changes in six populations of Tasmanian devils (*Sarcophilus harrisii*) before and after contracting a unique transmissible cancer known as devil facial tumour disease (DFTD). Using a RAD-capture approach, they identify 186 candidate genes that are enriched for cell cycling and immune response, indicating the genetic pathways that are important for viability selection after infection. They also use DNA sequence information to compare these contemporary signatures with molecular signatures of selection (i.e. historical selection over millions of years) to investigate if these loci have been the target of recurrent selection in devils; notably, there was limited overlap in regions under selection over these different evolutionary timescales.

### (d) Multivariate quantitative genetics: moving beyond the genotype-phenotype map.

Most research in wild quantitative genetics and genomics has focussed on individual traits within one species. However, trade-offs and genetic conflict of polygenic variation within and between individuals, sexes, and even species, are important components of life-history theory [58,59]. Genetic correlations between traits can constrain or facilitate selection, and may explain the paradox that evolutionary stasis is common despite evidence of frequent directional selection on heritable traits [60]. The ability of populations to respond to multivariate selection is well demonstrated by the Lande equation $\Delta z = G\beta$, where $\Delta z$ is the change in a vector of multivariate normal traits due to selection, **G** is the additive variance-covariance matrix (the structure of which can constrain evolution) and ***β*** is a vector of directional selection gradients [61,62]. The level of evolutionary constraint is also likely to be highly dependent on underlying genetic architectures [63]. Therefore, the integration of genomic data into multivariate models creates opportunities to investigate how genetic correlations can shape evolution from several different standpoints, as demonstrated by the submissions described here.

Several papers in this issue use QTL mapping of multiple traits to investigate whether correlated traits share a genetic basis. Coughlan *et al*. [37] demonstrated that some mid-to-large effect QTLs associated with correlated life history traits in *Mimulus guttatus* and *M. decorus* mapped to the same genomic regions. These QTLs explained life-history divergence between different parental populations, sometimes with the opposite effect than expected. In particular, delayed rhizome production over development in heterozygotes at a large-effect QTL may have implications for hybrid fitness in nature. Lin *et al*. [36] also identified overlapping QTL for correlated fitness and fitness-related traits in *Boechera stricta* from different parent populations & environments. The findings of both studies, particularly where opposing effects are observed, have implications for gene-flow, local adaptation and population divergence.

One phenomenon that can maintain genetic variation for fitness-related traits in natural populations is sexual antagonism, where selection favours different alleles within males and females [64]. Yet, the amount and prevalence of sexually antagonistic variation depends on the genetic architecture of dimorphism, which is not well understood [59,65]. Kollar et al. [44] used a multivariate approach to examine the genetic basis of sexual dismorphism in a scent-based fertilisation syndrome in the moss *Ceratodon purpureus*. They disentangled constraints shaping variation in volatile organic compounds (VOC) emissions, by analysing cross-sex genetic correlations within traits, sex-specific (co)variance between traits (i.e. the **G** matrix above), and cross-sex, cross-trait (co)variances. Their findings are highly trait dependent but are somewhat consistent with sex-specific architectures that may have evolved to resolve sexual conflict.

Reddiex and Chenoweth [43] also used the **G** matrix to investigate the maintenance of variation for phenotypic traits. They combined GWAS with multivariate models to investigate constraints on sexual selection on male cuticular hydrocarbons (CHCs) in wild-derived, genome-resequenced lines of *Drosophila serrata*. They estimated the genetic (co)variance matrix **G** in conjunction with multivariate SNP effects to detect genomic regions with the strongest multivariate associations. By investigating the (mis)alignment of **G** and the directional selection gradients *β*, they present evidence of multivariate genetic constraints on CHC evolution that are likely to be polygenic in nature.

Finally, it is long known that a trait of an individual can be influenced not only by its own genome, but also the genome of interacting individuals. This can arise through indirect genetic effects within the same species (e.g. maternal effects and/or behaviour of other related or unrelated individuals [66]), a phenomenon well-described in evolutionary genetics [67,68]. However, this can also arise through the genetic effects of other species; for example, some phenotypes measured in plants and animals are likely to be influenced by their microbiota, and therefore may map to and co-evolve with the genome of an entirely different species [69]. Yet, this phenomenon remains poorly understood. O'Brien et al. [40] used a simulation approach to investigate the genetic architecture and coevolution of a host-microbe mutualism under several evolutionary scenarios. In mutualisms, both species have fitness benefits, but this outcome can be influenced by conflict, i.e. when optimal trait values and/or strengths of selection are different for each species [70]. O'Brien *et al.* demonstrate that GWAS can map a joint trait to loci in multiple genomes, and they also show how fitness conflict and feedback loops can lead to variation in multi-genomic architectures. They show that fitnesses can also be positively correlated in spite of conflict, and that even when reaching a trait optimum (and often generating a strong mutualistic dependency), genetic variation and the evolutionary potential of a joint trait can be maintained.

## 4. Future outlook for the field.

The Wild Quantitative Genomic studies presented in this special issue demonstrate the promise of genomic technologies to better characterise quantitative genomic variation and to understand a number of outstanding evolutionary questions. Such studies are more accessible than ever before, and outcomes will have a wider impact in the fields of conservation genetics, molecular biology, and animal and plant breeding, among others. Yet, a number of immediate challenges remain, from the perennial issue of study power, to persistent structural barriers faced by researchers from the Global South and other historically-excluded groups. Here, we will discuss these promises and challenges in more detail, and consider the direction and legacy of wild quantitative genomics in the years to come.

**(a) Wild quantitative genomic studies are more feasible than ever before.**

Genomic technologies have made quantitative genetic studies more achievable in a wider range of systems. Pedigrees are no longer an explicit requirement for estimating heritabilities and breeding values in populations where enough genomic variation is captured [4,24,27,43,45,46]. Furthermore, the use of genomic estimated breeding values (GEBVs) overcomes numerous biases that arise when deriving pedigree-based estimates [56], particularly for individuals with fewer pedigree connections where environmental and genetic effects on phenotype are more likely to be confounded [71]. Genomic prediction also has the advantage of allowing the estimation of breeding values for individuals that have not been phenotyped [46,50]. The use of GEBVs is gaining traction in wild evolutionary studies [45,72,73] and will have important applications for investigating responses to selection over time, although there are limitations (see 4(c) below).

Although pedigrees are no longer required to estimate heritabilities and breeding values, long-term longitudinal data-sets are still vital for the field as they allow us to investigate temporal trends and microevolution [74]. Pedigrees remain an integral tool for many analyses, including: estimation of individual reproductive success [8]; validation of genomic heritability estimates [28]; determining if allele frequency changes are greater than those expected under genetic drift (i.e. simulation of Mendelian sampling, or "gene-dropping", of alleles through the pedigree) [75]; and for managing genetic diversity in threatened populations [76]. Indeed, methods have been developed that use genomic data to reconstruct pedigree relatedness between individuals, even in cases where connecting individuals are missing [77]; this can be conducted using relatively small numbers of molecular markers, which may be more cost-effective than larger-scale genomic studies. In addition, long-term datasets allow more fine-scale modelling of important covariates to quantify the effects and account for intrinsic and extrinsic variation within and between individuals, including common environment effects, indirect genetic effects, environment quality, population densities, and age [11]. Finally, long-term studies are important for building long-term partnerships with conservation practitioners and Indigenous guardians of vulnerable populations [76].

Quantitative genomics is also creating new opportunities to examine genetic architectures and evolution within multivariate frameworks. It is unlikely that quantitative traits evolve independently, but are rather constrained and/or coevolve with correlated traits. Identifying the genetic architecture of correlated, polygenic traits may help to elucidate the role of pleiotropy (i.e. genetic architecture shared between traits) and/or patterns of linkage disequilibrium between causal loci in driving trait correlations [78–80]. In addition, understanding the genetic architecture of cross-sex correlations and sex-by-locus interactions may explain the evolution of sexual dimorphism and the prevalence of sexual conflict in maintaining genetic variation and/or preventing phenotypes from reaching their sex-specific optima [43,44,64]. There is also more scope for investigating the genetics of interacting individuals and species [40], as well as potential interactions at the genotypic level with environmental heterogeneity [36,37].

**(b) The importance of polygenic variation is becoming clearer.**

Traditional quantitative genetics relies on the assumption of the infinitesimal model. However, with the advent of QTL mapping and GWAS, the focus shifted away from quantifying this polygenic component and instead investigating effects at loci of large effect [31,81]. These "low-hanging fruit" were easier to detect as they often explain a large proportion of trait variance, with empirical examples often shown to be maintained through stabilising selection [82–85], fluctuating selection [86] and/or local adaptation [87]. However, these observations may be the exception, rather than the rule [88]. Responses to selection on traits are often consistent with a polygenic architecture (i.e. the infinitesimal model), as responses to artificial directional selection are repeatedly shown to be steady over many generations [88,89]. The prevalence of polygenic architectures is also supported by GWAS results in model systems with very large sample sizes, where even if some of the genetic variance can be explained by candidate genes in core pathways, the majority of the genetic variance is explained by loci of small effect spread across the genome [90].

Fortunately, advances in quantitative genomics have led to better frameworks for determining the genetic architecture of traits, particularly in the absence of large effect loci. In previous QTL and GWAS studies, it was often not possible to distinguish if a lack of detectable loci was due to a trait being polygenic or due to a lack of power [30].

Now, models using genome partitioning [27] or genomic prediction [45] provide direct support for underlying polygenic architectures of many traits of interest, suggesting that the infinitesimal model may continue to be useful in many situations. Nevertheless, QTL and GWAS studies remain an essential tool in quantitative genomic studies. Major effect loci have been repeatedly uncovered for a range of quantitative traits, as demonstrated in this issue [35–37] and elsewhere [30], giving insight into underlying genetic mechanisms and their evolutionary dynamics [91]. In addition, GWAS summary statistics can be used to determine the proportion of loci that have non-zero effects on phenotypes, giving insight into the distribution of underlying effect sizes (and potentially the polygenicity) of traits [80,92]. However, studies investigating polygenic architectures are still subject to some limitations and issues with power, as outlined in section 4(c) below.

This special issue has also highlighted the utility of common population genetic approaches to characterise both simple and polygenic genomic signatures of selection in natural populations. Quantitative genetic studies are often focussed on specific (fitness-related) phenotypes, yet population genetic approaches can overcome the constraint of phenotype by focussing on variation in allele frequencies between different groups of individuals. This is demonstrated in this special issue by the use of selection component analyses by Kelly [47]. Such approaches can test for selection at different life stages where this information is available [93], providing a powerful framework for predicting allele frequency changes over time [94] and assessing the role of sexual conflict [95] or life history trade-offs in maintaining genetic variation [96]. Overall, both quantitative and population genomic approaches applied to individual-level genomic data can characterise simple and polygenic architectures of fitness and adaptation, to better understand and validate the underlying genetic mechanisms and pathways at play.

### (c) Wild quantitative genomic studies still face challenges.

A perennial and familiar issue is that wild quantitative genomic studies are often underpowered, particularly when they aim to determine the genetic architecture of traits [30]. The number of individuals that can be sampled, phenotyped, and genotyped are often limited (from 10s to 1,000s), which can lead to overestimation of effect sizes and/or false positive associations in QTL and GWA studies, a phenomenon known as the "Beavis effect" [97,98]. Another issue is that closely-related individuals may be more likely to share environments, which may upwardly bias heritability estimates if common environment effects are not or cannot be accounted for [11]. Furthermore, obtaining reliable genotypes with suitable marker density can itself be a challenge, especially in species without reference genomes or those with large genomes, or in cases where tissue samples cannot be easily prepared and preserved *in situ*. Estimation of trait heritabilities and breeding values using genomic approaches rely on markers being in linkage disequilibrium with causal variants, and so low marker densities could lead to underestimation of the genetic contribution to phenotype [25]. Reduced representation sequencing approaches have been popular for generating markers in less-tractable systems [99,100], and newer approaches involving imputation along with low-coverage short-read sequencing have potential to generate higher numbers of genotypes in non-model systems [101,102].

One route to understanding the effects of sample sizes and marker density on quantitative genomic analyses are simulation approaches that estimate and/or incorporate population demographic histories and markers densities [103], as well as replication and/or subsampling of data [104]. McGaugh et al. [46] in this issue also use simulation approaches to highlight the issue of power to accurately estimate GEBVs in outbred populations, demonstrating the requirement for appropriately-sized training datasets, high marker densities, moderate to high heritabilities, and high concordance between the modelled distribution of effect sizes and the true genetic architecture of focal traits. As a result, researchers should recognise the limitations of examining the microevolution of polygenic traits in a genomic prediction framework [46].

### (d) Quantitative genomics will be an important tool in conservation.

Advancements in the field of quantitative genomics can have important implications for conservation [105,106]. To date, most conservation genetics studies and management recommendations are based on the idea that maximising levels of genetic diversity is important for maintaining population viability in the face of changing environments [107]. While the importance of neutral genetic diversity in conservation has been debated [108,109], linking genomic variation to trait variation and adaptive potential will be useful for conservation goals. In turn, understanding the genetic basis of fitness-related traits and estimating adaptive potential in such populations can improve the population genetic theories underlying conservation genetic management and to inform conservation efforts in the future.

Studies in this special issue provide important lessons in this context. Duntsch et al. discuss how understanding the genetic architecture of ecologically-relevant traits in the hihi enables better predictions of adaptive potential and response to selection [27]. They caution that the adaptive potential of low heritability traits can be further constrained if those traits are polygenic, and how adaptive variation may be lost to drift faster than mutation can replenish it in populations with small effective population sizes. Stahlke et al. use a complementary approach to assess the genomic response to selection across different timescales, identifying potential loci that could be useful in future monitoring efforts for adaptation to DFT disease in Tasmanian devils [48]. McGaugh et al. also note that genomic prediction methods could inform captive breeding strategies to select individuals e.g. for resistance to disease and (theoretically) for accelerating responses to climate change [46]. While the potential of quantitative genomics to increase our ability to predict population trajectories and better inform conservation strategies is improving, we acknowledge that on-the-ground habitat protection and management is still of paramount importance in the protection of declining species.

**(e) Barriers to increasing diversity of researchers and research systems persist.**

Earlier, we mentioned that quantitative genetics can sometimes be a "tool of privilege". There is a large potential for this field to reach more diverse species and research questions, but there are still clear structural barriers to accessing genomics technologies, and for inclusion of researchers from historically excluded groups and the Global South [110]. The latter is a pervasive problem in ecology and evolution [111–115]. If both barriers are unaddressed, we will miss critical insights from species in the most biodiverse and vulnerable habitats and, more importantly, novel research directions, approaches, and perspectives from researchers from diverse backgrounds [116]. As a field, it is imperative that we collectively dismantle these structural barriers through actions such as: fixing disparities in access to funding [117], publishing [118,119], and networking opportunities [120]; critically examining our tendency to associate with similar people and instead fostering more inclusive recruitment, authorship, and collaboration practices [112]; and reducing helicopter science by building meaningful relationships with indigenous guardians and local communities to ensure equitable cogeneration of knowledge and sharing of benefits [111].

We also wish to note that this special issue was compiled during the first years of the COVID-19 pandemic. Whilst this was a difficult period for the life and work of many, it is now well acknowledged that the pandemic has exacerbated systemic inequalities related to race and gender in science [121], and is therefore likely to have had a similar impact within the wild quantitative genomics field.

## 5. Conclusion

The field of wild quantitative genomics has a rich history of adapting classic approaches for investigating genetic variation of fitness-related traits, whilst dealing with individual and environmental heterogeneity and challenges of statistical power. Genomic technologies have widened the questions and opportunities to better understand the variation we see in the world around us. The lessons we learn in wild populations inform our basic understanding of the drivers and evolutionary dynamics of quantitative variation in any population. We hope that this issue shows the continued promise of these approaches for better understanding the natural world, as well as providing a solid foundation of knowledge that can inform applied research in conservation, animal and plant breeding, and medical and human genetics.

### Glossary

- **Adaptive potential:** the ability of a population to respond to selection via adaptation, typically measured as the **additive genetic variance** for fitness.
- **Epigenetics:** changes in the genome, such as DNA methylation, that can have phenotypic effects but do not involve changes in nucleotide sequences.
- **Fitness:** the contribution of an individual to the next generation or reproductive pool. Fitness is often estimated using fitness-related traits or components, such as survival, fecundity, or growth rate.
- **Genetic architecture:** the number, effect sizes, frequencies and/or interactions of genes that underlie a trait. Genetic architecture can be described as **monogenic** or **Mendelian** if variation at one gene shapes trait variation, and **polygenic** if variation at many genes shapes trait variation.

- **Genetic correlation:** the proportion of genetic variance that is shared between traits. These are estimated in **multivariate models**, which simultaneously analyses two or more phenotypes. Genetic correlations can be caused by **linkage disequilibrium** between **QTLs** for two traits, or **pleiotropy** where the same **QTLs** affect multiple traits.
- **Genomic prediction**: a method using genome-wide molecular SNP markers to predict **genomic estimated breeding values (GEBVs)** for quantitative traits within individuals. It can also be used to estimate the effect sizes of individual SNP markers.
- **Genomic relatedness matrix (GRM)**: a covariance matrix calculated based on the similarity of SNP genotypes between individuals. This can be used to calculate the narrow-sense **heritability** or to account for phenotypic variation shaped by background genetics and shared environment in a **GWAS**.
- **Genome-wide association study (GWAS):** modelling the association between genome-wide SNP loci and a phenotype, used to identify regions of the genome associated with trait variation.
- **Heritability:** the proportion of trait variance that is contributed by genetic variation. The "narrow-sense" heritability is trait variance that is attributable to additive effects of genetic variants ('**additive genetic variance**).
- **Inbreeding depression:** the decrease in fitness associated with increasing homozygosity in the genome, most often due to matings between related individuals.
- **Indirect genetic effects:** the variance of a trait that is contributed by additive genetic effects in other conspecific individuals, e.g. maternal genetic effects.
- **Life history trait:** a trait that is related to the pattern of survival and reproduction events over an organism's lifespan.
- **Linkage disequilibrium:** the non-random association of alleles at a pair of loci.
- **Quantitative genetics:** the study of the genetic basis of continuous/complex/quantitative traits.
- **Quantitative trait locus (QTL):** a locus associated with variation in a quantitative trait. Can be identified through QTL-mapping or GWAS. QTLs may be: the **causal variant** that directly contributes to trait variation; a region containing that variant; or neutral variation that is in **linkage disequilibrium** with a causal variant.

## Data accessibility

This article has no additional data.

## Authors' contributions

All authors wrote and edited the manuscript, and approved the final version.

## Competing interests

We declare we have no competing interests.

## Funding

## Acknowledgements

We thank Shalene Singh-Shepherd for her patience and understanding on compiling this issue during the first two years of the COVID-19 pandemic and periods of family leave by the authors. We thank Gary Carvalho, who acted as Senior Editor for all of our papers. We thank Anna Santure, Loeske Kruuk, and two anonymous reviewers who made insightful suggestions on this Editorial piece. Finally, we thank all of the authors and reviewers who contributed to this special issue, particularly given the personal and professional challenges of the pandemic.